\documentclass{elsart}
\usepackage{amsfonts}
\usepackage{amsmath}

\input{tcilatex}

\begin{document}

\begin{frontmatter}%

\title{Quantum Theory of Irreversibility}%

\author{A. Pérez-Madrid}%
\footnote{%
E-mail address: agustiperezmadrid@ub.edu}

\address{Departament de Física Fonamental. Facultat de Física. Universitat de Barcelona. Diagonal 647, 08028 Barcelona. Spain}%

\begin{abstract}
A generalization of the Gibbs-von Neumann relative entropy is proposed based
on the quantum BBGKY [Bogolyubov-Born-Green-Kirkwood-Yvon] hierarchy as the
nonequilibrium entropy for an N-body system. By using \ a generalization of
the Liouville-von Neumann equation describing the evolution of a density
superoperator, it is demonstrated that the entropy production for an
isolated system is non-negative, which provides an arrow of time. Moreover,
following the procedure of non-equilibrium thermodynamics \ a\ \ master
matrix is introduced for which a microscopic expression is obtained. Then,
the quantum Boltzmann equation is derived in terms of a transition
superoperator related to that master matrix. 
\end{abstract}%

\begin{keyword}
quantum statistical mechanics, nonequilibrium and irreversible
thermodynamics, kinetic and transport theory of gases 
\end{keyword}%

\end{frontmatter}%

\section{Introduction}

According to the mechanicistic interpretation of the physical world, the
basic laws of nature are deterministic and time reversible. However, at the
macroscopic level we observe irreversible processes related to energy
degradation which generate entropy. How do we reconcile the `spontaneous
production of entropy' with the time reversibility of the microscopic
equations of motion?. This is a problem not solved in a clear and definitive
way yet, thus it still constitutes an open problem. Attempts to address this
have been made based on coarse-graining procedures \cite{dorfman}-\cite%
{rondoni}. In a previous work \cite{agusti} we showed that in the framework
of the classical BBGKY- hierarchy which explicitly incorporates the
correlation between all the particle clusters in the system, macroscopic
irreversibility arises without coarse-grainig procedures. All that is needed
is a coherent definition of the entropy given through a generalization of
the Gibbs relative entropy in terms of the distribution vector in addition
to the generalized Liouville equation. Based on the fact that the
generalized Liouville equation introduces a trace preserving transformation
in the set of distribution vectors, we obtained a non-negative entropy
production for the isolated system which manifests the existence of a
classical time arrow by appealing to the underlying microscopic dynamics
only.

Here we develop the quantum version of this previous analysis. The quantum
counterpart of the distribution vector is the density superoperator evolving
according to the generalized Liouville-von Neumann equation. Moreover, we
propose a generalization of the Gibbs-von Neumann relative entropy in terms
of the density superoperator, in the spirit of the generalized Gibbs
relative entropy postulate. As in the classical approach, here, the
generalized Liouville-von Neumann equation introduces a trace preserving
transformation aswell. From this we obtain a non-negative entropy production
for the isolated system, providing an arrow of time and giving a microscopic
basis to the second law of \ Thermodynamics. After obtaining that, through a
thermodynamic analysis, the quantum Boltzmann equation follows in a natural
way.

We have structured the paper as follows. In section 2, we introduce the
Hamiltonian dynamic of the N-body system obtaining the generalized
Liouville-von Neumann equation. In section 3, we compute the entropy
production. Section 4 is devoted to the non-Equilibrium Thermodynamic
analysis of the system obtaining the quantum Boltzmann equation. Finally in
section 5, we emphasize our main conclusions.

\section{Hamiltonian Dynamics}

Let's consider a dynamical system of N identical particles whose Hamiltonian
is given by the N-particle Hamiltonian%
\begin{equation}
H^{(N)}=\sum_{j=1}^{N}H(j)+\frac{1}{2}\sum_{j\neq k=1}^{N}\phi\left(
j,k\right) \text{ ,}  \label{hamiltonian}
\end{equation}
where $H(j)$ denotes the individual energy of the $j$-th particle and $%
\phi\left( j,k\right) $ the interaction energy between the $j$-th and $k$-th
particles. The state of the system is completely specified at a given time
by the N-particle density operator $D^{(N)}(1,...,N)$ which evolves
according to the Liouville-von Neumann equation \cite{bogoliuvov}

\begin{equation}
i\hbar \frac{\partial }{\partial t}D^{(N)}=\left[ H^{(N)},D^{(N)}\right] 
\text{ \ .}  \label{von Neumann}
\end{equation}%
In the Liouville representation, where the density operator is represented
as a vector in the Liouville space \cite{balazs}-\cite{rau}, the
Liouville-von Neumann equation becomes 
\begin{equation}
i\hbar \frac{\partial }{\partial t}D_{\alpha \beta }^{(N)}=L_{\alpha \beta
,\gamma \delta }D_{\gamma \delta }^{(N)}\text{ ,}  \label{liouville}
\end{equation}%
where repeated Greek indices mean summation and the Liouvillian $L$ is
defined through%
\begin{equation}
L_{\alpha \beta ,\gamma \delta }=\left( H_{\alpha \delta }^{(N)}\delta
_{\beta \gamma }-H_{\gamma \beta }^{(N)}\delta _{\alpha \delta }\right) 
\text{ \ .}  \label{liouvillian}
\end{equation}%
However, a strictly equivalent alternative description of the state of the
system can be given in terms of the set of n-particle reduced density
operators \cite{bogoliuvov}, \cite{dufty} 
\begin{equation}
\mathcal{D}\equiv \left\{ D^{(0)},D^{(1)},......,D^{(N)}\right\} \text{ \ ,}
\label{superoperator}
\end{equation}%
where $\mathcal{D}$ constitutes a density superoperator whose components are
the n-particle reduced density operators $D^{(n)}(1,....,n)$ defined through 
\begin{equation}
D^{(n)}(1,....,n)=\frac{N!}{(N-n)!}\underset{(n+1,...,N)}{\text{Tr}}%
D^{(N)}(1,....,N)\text{ \ }  \label{reduced density}
\end{equation}%
and $D_{\alpha \beta }^{(0)}=\delta _{\alpha \beta }$ (\textit{i.e.} $%
D^{(0)} $ coincides with the unit matrix). These operators satisfy%
\begin{equation}
\underset{_{\left( 1,...,n\right) }}{\text{Tr}}\left\{
D^{(n)}(1,....,n)\right\} \text{ =}\frac{N!}{(N-n)!}\text{ \ .}
\label{normalization}
\end{equation}

\bigskip The superoperator $\mathcal{D}$ can be represented alternatively as
a $(N+1)\times(N+1)$ block diagonal matrix in an abstract space%
\begin{equation}
\mathcal{D=}\dsum \limits_{n=0}^{N}\frac{1}{n!}\mathcal{I}_{n}\otimes D^{(n)}%
\text{ ,}  \label{representation}
\end{equation}
where the $(N+1)\times(N+1)$ matrices $\mathcal{I}_{n}$ are characterized
for having zero everywhere except for the n-th row and the n-th column and
satisfying%
\begin{equation}
\mathcal{I}_{n}\odot\mathcal{I}_{p}=\delta_{np}  \label{inner}
\end{equation}
which defines the inner product in this abstract space. Moreover, in view of
\ Eq. (\ref{normalization})%
\begin{equation}
\text{Tr }\left\{ \mathcal{D}\right\} =\dsum \limits_{n=0}^{N}\frac{1}{n!}%
\underset{(1,...,n)}{\text{Tr}}\left\{ D^{(n)}\right\} =2^{N}\text{ .}
\label{trace}
\end{equation}

On the other hand, the dynamics of the $D^{(n)}$ follows after performing
the operation $\underset{(n+1,...,N)}{\text{Tr}}$ on both sides of Eq. (2).
Thus, we obtain the so-called quantum BBGKY-hierarchy of equations which can
be written in a compact way as a generalized Liouville equation%
\begin{equation}
i\hbar\frac{\partial}{\partial t}\mathcal{D}(t)=\mathcal{L}\odot \mathcal{D}%
(t)\text{ \ .}  \label{generalized liouville}
\end{equation}
At this point, it is convenient to introduce the projection operators $%
\mathcal{P}$ and $\mathcal{Q}$, its complement with respect to the identity $%
\left( \text{\textit{i.e.}, \textit{\ }}\mathcal{PL}+\mathcal{QL}=\mathcal{L}%
\right) $, which give the diagonal and non-diagonal part of the generalized
Liouvillian $\mathcal{L}$, respectively. This projection operators are
defined according to 
\begin{equation}
\mathcal{PL}=\sum_{n=0}^{N}\sum_{p=0}^{N}\delta_{pn}\mathcal{I}_{n}\otimes%
\mathcal{I}_{n}\otimes L^{\left( n\right) }\text{,}  \label{diagonal}
\end{equation}
where $L^{\left( n\right) }$ is the n-particle Liouvillian 
\begin{equation}
L_{\alpha\beta,\gamma\delta}^{\left( n\right) }=\left( H_{\alpha\delta
}^{(n)}\delta_{\beta\gamma}-H_{\gamma\beta}^{(n)}\delta_{\alpha\delta}\right)
\label{reducedliouvillian}
\end{equation}
and $H^{(n)}$ is the n-particle Hamiltonian. Thus, it becomes 
\begin{equation}
\mathcal{PL}\odot\mathcal{D=}\dsum \limits_{n=0}^{N}\frac{1}{n!}\mathcal{I}%
_{n}\otimes\left( L^{\left( n\right) }\odot D^{(n)}\right) \text{ .}
\label{test}
\end{equation}
In addition, when the inner product $\odot$ involves two densities operators 
$U$ and $V$, \ it should be understood as 
\begin{equation}
A\odot B=A_{\alpha\beta}B_{\beta\alpha}\text{ \ .}  \label{inner product}
\end{equation}
On the other hand, 
\begin{equation}
\mathcal{QL}=\sum_{n=0}^{N}\mathcal{I}_{n}\otimes\sum_{p=0}^{N}\delta
_{p(n+1)}\underset{p}{\text{Tr}}\text{ \ }\left\{ \mathcal{I}%
_{p}\otimes\Delta^{(p)}\right\} \text{,}  \label{nondiagonal}
\end{equation}
where we have defined%
\begin{equation}
\Delta_{\alpha\beta,\gamma\delta}^{(p)}=\left( F_{\alpha\delta}^{\left(
p\right) }\delta_{\beta\gamma}-F_{\gamma\beta}^{(p)}\delta_{\alpha\delta
}\right)  \label{antisym}
\end{equation}
and 
\begin{equation}
F^{\left( p\right) }(1,....,p)=\dsum \limits_{j=1}^{p-1}\phi\left(
j,p\right) \text{ \ .}  \label{interaction}
\end{equation}
Therefore,%
\begin{equation}
\mathcal{QL}\odot\mathcal{D=}\dsum \limits_{n=0}^{N}\frac{1}{n!}\mathcal{I}%
_{n}\otimes\underset{(n+1)}{\text{Tr}}\left\{ \Delta^{(n+1)}\odot
D^{(n+1)}\right\} \text{ \ .}  \label{test2}
\end{equation}
Hence, Eq. (\ref{generalized liouville}) can be rewritten as 
\begin{equation}
i\hbar\frac{\partial}{\partial t}\mathcal{D}(t)-\mathcal{PL}\odot \mathcal{D}%
=\mathcal{QL}\odot\mathcal{D}\text{ .}  \label{generalized2}
\end{equation}
According to Eqs. (\ref{diagonal})-(\ref{interaction}), one of the
components of \ Eq. (\ref{generalized2}) will read 
\begin{align}
& i\hbar\frac{\partial}{\partial t}D_{\alpha\beta}^{(n)}-\left(
H_{\alpha\delta}^{(n)}D_{\beta\delta}^{(n)}-H_{\gamma\beta}^{(n)}D_{\alpha%
\gamma}^{(n)}\right)  \notag \\
& =\underset{(n+1)}{\text{Tr}}\left\{ F_{\alpha\delta}^{\left( n+1\right)
}D_{\beta\delta}^{(n+1)}-F_{\gamma\beta}^{(n+1)}D_{\alpha\gamma}^{(n+1)}%
\right\} \text{ \ .}  \label{hierarchy}
\end{align}
Without the right hand side term in this equation, the evolution of $D^{(n)}$
would be unitary. In the next section we will show that irreversibility is
manifested in the dynamics of the system when the adequate description, 
\textit{i.e. }in terms of the set $\mathcal{D}$ of all the n-particle
reduced density operators is used.

\section{\protect\bigskip Non-equilibrium entropy and irreversibility}

As the expression for the non-equilibrium entropy we propose%
\begin{align}
S & =k_{B}\text{Tr}\left\{ \mathcal{D}\left( \ln\mathcal{D}_{o}\mathcal{-}\ln%
\mathcal{D}\right) \right\} +S_{o}  \notag \\
& =k_{B}\sum_{n=0}^{N}\frac{1}{n!}\mathcal{I}_{n}\otimes D^{(n)}\odot
\sum_{p=0}^{N}\mathcal{I}_{p}\otimes\left( \ln D_{o}^{(p)}-\ln
D^{(p)}\right) +S_{o}  \notag \\
& =k_{B}\sum_{n=0}^{N}\frac{1}{n!}D^{(n)}\odot\left( \ln D_{o}^{(n)}-\ln
D^{(n)}\right) \;+S_{o}\text{ ,}  \label{gibbs}
\end{align}
a functional of $D^{(n)}$, generalizing the Gibbs-von Neumann entropy
postulate \cite{degroot}, \cite{vankampen}, based on the fact that the
superoperator $\mathcal{D}$ determines the state of the system. This entropy
is analogue to the relative or Kullback entropy \cite{schlogl}, \cite{wherl}%
. Here, $k_{B}$ is the Boltzmann constant, $S_{o}$ the equilibrium entropy
and $\mathcal{D}_{o}$ is assumed to be the equilibrium value of $\mathcal{D}$
which corresponds with $S_{o}$, satisfying $\mathcal{LD}_{o}=0$. It is known
that $S$ is a concave functional of $\mathcal{D}$, and will never decrease
under a linear stochastic transformation like the one induced by Eq. (\ref%
{generalized liouville}) \cite{lieb}-\cite{wherl} 
\begin{equation}
\mathcal{D}(t+\tau)=(\mathcal{U}+\tau\mathcal{L})\odot\mathcal{D}(t)\text{ ,}
\label{stochastic}
\end{equation}
with $\mathcal{U}$ being the unit superoperator%
\begin{equation}
\mathcal{U}\text{ }=\sum_{n=0}^{N}\mathcal{I}_{n}\otimes\mathcal{I}%
_{n}\otimes U^{(n\mid n)}  \label{superunit}
\end{equation}
and 
\begin{equation}
U^{(n\mid n)}=\dsum
\limits_{\alpha\beta,\gamma\delta}I_{\alpha\beta,\gamma\delta}\left\vert
\alpha\right\rangle \left\langle \beta\right\vert \otimes\left\vert
\gamma\right\rangle \left\langle \delta\right\vert \text{ \ ,}  \label{unit}
\end{equation}
where $I_{\alpha\beta,\gamma\delta}=\delta_{\gamma\beta}\delta_{\alpha%
\delta} $ is the unit tetradic \cite{Zwanzig}. Therefore, the entropy we
propose is coherent with the second law according to which $S$ increases in
irreversible processes such as the relaxation to equilibrium from an
initially non-equilibrium state. To prove this we compute the rate of change
of $S$ which can be obtained by differentiating Eq. (\ref{gibbs}) with the
help of Eq. (\ref{generalized liouville}) 
\begin{align}
\frac{dS}{dt} & =k_{B}\text{Tr}\left\{ \frac{\partial\mathcal{D}}{\partial t}%
\left( \ln\mathcal{D}_{o}\mathcal{-}\ln\mathcal{D}\right) \right\}
\label{entropyproduction} \\
& =\sigma_{1}+\sigma_{2}\text{ }\geq0\text{ , }  \notag
\end{align}
where%
\begin{equation}
\sigma_{1}=-i\hbar^{-1}k_{B}\text{Tr}\left\{ \mathcal{PL}\odot\mathcal{D}%
\left( \ln\mathcal{D}_{o}\mathcal{-}\ln\mathcal{D}\right) \right\} \text{ ,}
\label{sigma1}
\end{equation}
and 
\begin{equation}
\sigma_{2}=-i\hbar^{-1}k_{B}\text{Tr}\left\{ \mathcal{QL}\odot\mathcal{D}%
\left( \ln\mathcal{D}_{o}\mathcal{-}\ln\mathcal{D}\right) \right\} \text{ .}
\label{sigma2}
\end{equation}
Eq. (\ref{entropyproduction}) constitutes the entropy production
corresponding to the relaxation passing from a non-equilibrium state to
equilibrium. By using Eqs. (\ref{diagonal})-(\ref{test2}) in addition to the
cyclic invariance property of the trace, one has%
\begin{align}
\sigma_{1} & =-i\hbar^{-1}k_{B}\sum_{n=0}^{N}\frac{1}{n!}L^{\left( n\right)
}\odot D^{(n)}\odot\left( \ln D_{o}^{(n)}-\ln D^{(n)}\right)  \notag \\
& =-i\hbar^{-1}k_{B}\sum_{n=0}^{N}\frac{1}{n!}\left[ H^{\left( n\right)
},D^{(n)}\right] \odot\left( \ln D_{o}^{(n)}-\ln D^{(n)}\right)  \notag \\
& =i\hbar^{-1}k_{B}\sum_{n=0}^{N}\frac{1}{n!}\left\{ H^{\left( n\right)
}\odot\left[ \ln D_{o}^{(n)},D^{(n)}\right] \right\}  \label{sigma11} \\
& =i\hbar^{-1}k_{B}\sum_{n=0}^{N}\frac{1}{n!}\left[ H^{\left( n\right) },\ln
D_{o}^{(n)}\right] \odot D^{(n)}  \notag
\end{align}
and%
\begin{align}
\sigma_{2} & =-i\hbar^{-1}k_{B}\sum_{n=0}^{N}\frac{1}{n!}\underset{(n+1)}{%
\text{Tr}}\left\{ \Delta^{(n+1)}\odot D^{(n+1)}\right\} \odot\left( \ln
D_{o}^{(n)}-\ln D^{(n)}\right)  \notag \\
& =-i\hbar^{-1}k_{B}\sum_{n=0}^{N}\frac{1}{n!}\underset{(n+1)}{\text{Tr}}%
\left\{ \left[ F^{\left( n+1\right) },D^{(n+1)}\right] \right\} \odot\left(
\ln D_{o}^{(n)}-\ln D^{(n)}\right)  \notag \\
& =i\hbar^{-1}k_{B}\sum_{n=0}^{N}\frac{1}{n!}\underset{(n+1)}{\text{Tr}}%
\left\{ F^{\left( n+1\right) }\odot\left[ \left( \ln D_{o}^{(n)}-\ln
D^{(n)}\right) ,D^{(n+1)}\right] \right\} \text{ \ .}
\label{entropyproduction2}
\end{align}
These are zero at equilibrium when $D^{(n)}=D_{o}^{(n)}$, but in any other
case these are not necessarily zero because neither the density $D^{(n+1)}$
commutes with $D^{(n)}$ and $D_{o}^{(n)}$ nor $D_{o}^{(n)}$ commutes with $%
H^{\left( n\right) }$, \ in general. Hence, according to our previous
discussion related to the concave character of the entropy $S$ and its
consequences, $\sigma$ should be non negative. This provides a time arrow
inherent to the irreversible macroscopic manifestation of the dynamics of
the system.

As known due to the unitary character of the time-evolution operator related
to the Liouville-von Neumann equation giving the dynamics of the N-particle
density operator, the von Neumann entropy is a constant of motion.
Nevertheless, as we have seen here, this is not the case for the generalized
Gibbs-von Neumann entropy, based on the fact that the dynamics of the
density superoperator given through the genaralized Liouville-von Neumann
equation is not unitary. In addition, as compared to other definitions of
entropy, Eq. (\ref{gibbs}) does not contain any coarse-grainig. In the light
of these facts not contained in other definitions of the entropy in the
literature the strength of our definition (\ref{gibbs}) resides.

\section{Non-equilibrium thermodynamics analysis}

Here we proceed as in Ref. \cite{mazur}, which constitutes the standard way
of Non-equilibrium thermodynamics. Thus, for small deviations from
equilibrium we assume that $\mathcal{D}$ and $\mathcal{D}_{o}$ are diagonal
in a common basis. According to this, the entropy (\ref{gibbs}) becomes up
to second order 
\begin{equation}
S=-\frac{1}{2}k_{B}\text{Tr}\left\{ \delta\mathcal{DD}_{o}^{-1}\mathcal{%
\delta D}\right\} +S_{o}\text{ ,}  \label{secondorder}
\end{equation}
where $\delta\mathcal{D}=\mathcal{D}-\mathcal{D}_{o}$. Therefore, the
entropy production is 
\begin{align}
\sigma & =\frac{dS}{dt}=-\frac{1}{2}k_{B}\text{Tr}\left\{ \frac {\partial%
\mathcal{D}}{\partial t}\left( \mathcal{D}_{o}^{-1}\delta \mathcal{D+}\delta%
\mathcal{DD}_{o}^{-1}\right) \right\}  \label{newproduction} \\
& =\text{Tr}\left\{ \frac{\partial\mathcal{D}}{\partial t}\mathcal{X}%
\right\} \text{ \ ,}  \notag
\end{align}
where the thermodynamic force%
\begin{equation}
\mathcal{X}\equiv\delta S/\delta\mathcal{D}=-\frac{1}{2}k_{B}\left( \mathcal{%
D}_{o}^{-1}\delta\mathcal{D+}\delta\mathcal{DD}_{o}^{-1}\right)
\label{force}
\end{equation}
is conjugated to the flux $\partial\mathcal{D}/\partial t$. From Eq. (\ref%
{newproduction}) the phenomenological relation 
\begin{equation}
\frac{\partial\mathcal{D}}{\partial t}=\mathcal{R}\odot\mathcal{X}
\label{linearelation}
\end{equation}
is inferred, where%
\begin{equation}
\mathcal{R}=\sum_{n=0}^{N}\sum_{p=0}^{N}\mathcal{I}_{n}\otimes\mathcal{I}%
_{p}\otimes R^{(n\mid p)}\text{ }  \label{m}
\end{equation}
is a phenomenological supermatrix which due to the Hermitian character of
the force $\mathcal{X}$ , we choose Hermitian on the right-hand side entry.
This implies that%
\begin{equation}
R_{\alpha\beta,\gamma\delta}^{(n\mid
p)}=R_{\alpha\beta,\delta\gamma}^{\ast(n\mid p)}\text{ ,}
\label{hermiticity1}
\end{equation}
where the superscript * stands for the complex conjugate. Using this
property, one can replace $\mathcal{X}$ \ in Eq. (\ref{linearelation}) by 
\begin{equation}
\mathcal{X}=-k_{B}\mathcal{D}_{o}^{-1}\delta\mathcal{D\ =}\dsum
\limits_{n=0}^{N}\frac{1}{n!}\mathcal{I}_{n}\otimes X^{(n)}\text{,}
\label{nonhermitian}
\end{equation}
a nonHermitian form. Furthermore, due to the fact that $\mathcal{D}$ is
normalized, from Eq. (\ref{linearelation}) we infer that Tr$(\mathcal{R}\odot%
\mathcal{X})=0$, which leads to the following constraints%
\begin{equation}
\dsum \limits_{\alpha}R_{\alpha\alpha,\gamma\delta}^{(n\mid p)}=0\text{ .}
\label{constrain}
\end{equation}
By substituting the phenomenological relation (\ref{linearelation}) into Eq.
(\ref{newproduction}) we reach the expression%
\begin{equation}
i\hbar^{-1}\sum_{n=0}^{N}\frac{1}{n!}\underset{(n+1)}{\text{Tr}}\left\{ %
\left[ X^{(n)},D^{(n+1)}\right] \odot F^{\left( n+1\right) }\right\}
=\sum_{n=0}^{N}\sum_{p=0}^{N}X^{(n)}\odot R^{(n\mid p)}\odot X^{(p)}\text{ ,}
\label{Green-Kubo}
\end{equation}
a kind of Green-Kubo relation, also obtained by using Eqs. (\ref{sigma11})
and (\ref{entropyproduction2}) in the near-equilibrium approximation.

Additionally, $\mathcal{R}$ should be Hermitian as predicted by the Onsager
symmetry relations, which reflects in%
\begin{equation}
R_{\alpha\beta,\gamma\delta}^{(n\mid
p)}=R_{\delta\gamma,\beta\alpha}^{\ast(p\mid n)}\text{ .}
\label{hermiticity}
\end{equation}
This leads to the constraint%
\begin{equation}
\dsum \limits_{\alpha}R_{\gamma\delta,\alpha\alpha}^{(n\mid p)}=0\text{ ,}
\label{constrain2}
\end{equation}
and to the hermiticity in the left-hand side entry%
\begin{equation}
R_{\alpha\beta,\gamma\delta}^{(n\mid
p)}=R_{\beta\alpha,\gamma\delta}^{\ast(n\mid p)}\text{ .}
\label{hermiticity3}
\end{equation}

With Eq. (\ref{nonhermitian}), Eq. (\ref{linearelation}) may be written in
the form%
\begin{equation}
\frac{\partial\mathcal{D}}{\partial t}=-\mathcal{M}\odot\delta\mathcal{D}%
\text{ ,}  \label{liearrelation2}
\end{equation}
where the master matrix is defined by%
\begin{equation}
\mathcal{M\equiv}\text{ }-k_{B}\mathcal{RD}_{o}^{-1}\text{ .}
\label{mastermatrix}
\end{equation}
According to the Onsager relations, it follows that the master matrix
satisfies%
\begin{equation}
\mathcal{MD}_{o}=-k_{B}\mathcal{R}\text{ }=-k_{B}\mathcal{R}^{\dag }=%
\mathcal{D}_{o}\mathcal{M}^{\dag}\text{ ,}  \label{detailedbalance}
\end{equation}
where $\dagger$ refers to the Hermitian conjugate. This results in the
detailed balance for transition probabilities, and vice versa, when those
are introduced.

By using the constraint Eq. (\ref{constrain2}) one has 
\begin{equation}
\mathcal{M\odot D}_{o}=-k_{B}\mathcal{RD}_{o}^{-1}\mathcal{\odot D}%
_{o}=-k_{B}\sum_{n=0}^{N}\sum_{p=0}^{N}\mathcal{I}_{n}\otimes R^{(n\mid
p)}\odot D^{(0)}=0\text{ }  \label{identity1}
\end{equation}%
which proves that the equilibrium density superoperator is a right
eigenfunction of the master supermatrix with an eigenvalue zero. Here, we
have used the fact that 
\begin{equation}
\mathcal{I}_{n}\mathcal{I}_{p}=\mathcal{I}_{n}\delta _{np}\text{ .}
\label{identity2}
\end{equation}%
So that, Eq. (\ref{liearrelation2}) can be rewritten as%
\begin{equation}
\frac{\partial \mathcal{D}}{\partial t}=-\mathcal{M}\odot \mathcal{D}\text{ .%
}  \label{linearrelation3}
\end{equation}%
Now, by substitution of the master equation (\ref{linearrelation3}) into the
generalized Liouville equation (\ref{generalized2}) we obtain%
\begin{equation}
\frac{\partial }{\partial t}\mathcal{D}(t)+i\hbar ^{-1}\mathcal{PL}\odot 
\mathcal{D}(t)=-\mathcal{QM}\odot \mathcal{D}(t)\text{ ,}
\label{generalized liouville2}
\end{equation}%
where $\mathcal{QM}$ should have the same structure as $\mathcal{QL}$, thus%
\begin{equation}
\mathcal{QM}=\sum_{n,p=0}^{N}p!\delta _{p(n+1)}\mathcal{I}_{n}\otimes 
\underset{\left( n+1\right) }{\text{Tr}}\text{ \ }\left\{ \mathcal{I}%
_{p}\otimes M^{(n+1\mid p)}\right\} \text{ }  \label{qm}
\end{equation}%
and 
\begin{equation}
\mathcal{Q\mathcal{M}}\odot \mathcal{D=}\dsum\limits_{n,p=0}^{N}\delta
_{p(n+1)}\mathcal{I}_{n}\otimes \underset{(n+1)}{\text{Tr}}\left\{
M^{(n+1\mid p)}\odot D^{(p)}\right\} \text{ .}  \label{qm2}
\end{equation}%
This last equation allows us to write%
\begin{gather}
\frac{\partial }{\partial t}D_{\alpha \beta }^{(n)}+i\hbar ^{-1}\left(
H_{\alpha \delta }^{(n)}D_{\beta \delta }^{(n)}-H_{\gamma \beta
}^{(n)}D_{\alpha \gamma }^{(n)}\right) =  \label{master} \\
-\dsum\limits_{p}\delta _{p(n+1)}\underset{(n+1)}{\text{Tr}}\left( M_{\alpha
\beta ,\gamma \delta }^{(n+1\mid p)}D_{\delta \gamma }^{(p)}\right) \text{ ,}
\notag
\end{gather}%
equivalent to Eq. (\ref{hierarchy}).

Here, Eq. (\ref{master}) can be put into a more common form by introducing
the transition superoperator $W_{\alpha \beta ,\gamma \delta }^{(n\mid p)}$
defined through%
\begin{equation}
W_{\alpha \beta ,\gamma \delta }^{(n\mid p)}=-M_{\gamma \delta ,\alpha \beta
}^{(p\mid n)}+\Lambda _{\alpha \beta ,\gamma \delta }^{(n\mid p)}\text{ ,}
\label{transitionmatrix}
\end{equation}%
where%
\begin{equation}
\Lambda _{\alpha \beta ,\gamma \delta }^{(n\mid p)}=I_{\alpha \beta ,\gamma
\delta }\Psi _{\gamma \delta }^{(n)}\delta _{np}\text{ .}  \label{auxiliary}
\end{equation}%
The auxiliary matrix $\Psi _{\gamma \delta }^{(n)}$ is not arbitrary,
satisfying 
\begin{equation}
\sum_{\alpha }W_{\alpha \alpha ,\gamma \delta }^{(n\mid p)}=\delta _{\gamma
\delta }\Psi _{\gamma \gamma }^{(n)}\delta _{np}\text{ ,}  \label{auxilary1}
\end{equation}%
in view of Eq. (\ref{constrain}). This means that $\sum_{\alpha }W_{\alpha
\alpha ,\gamma \delta }^{(n\mid p)}$ is a diagonal matrix. In order to
satisfy detailed balance, $\Lambda _{\alpha \beta ,\gamma \delta }^{(n\mid
p)}$ should be Hermitian. Hence, with Eq. (\ref{detailedbalance}) one has%
\begin{equation}
D_{o,\delta \gamma }^{(p)}W_{\gamma \delta ,\alpha \beta }^{(p\mid
n)}=W_{\beta \alpha ,\delta \gamma }^{(n\mid p)}D_{o,\gamma \delta }^{(p)}%
\text{ \ }  \label{detailedbalance2}
\end{equation}%
constituting the expression of the detailed balance principle. Thus,
projecting Eq. (\ref{master}) onto its diagonal part by means of the
tetradic operator $P_{ij,kl}=\delta _{ij}\delta _{ik}\delta _{jl\text{ }}$ 
\cite{Zwanzig},%
\begin{gather}
P\odot \left( \frac{\partial }{\partial t}D^{(n)}+i\hbar ^{-1}\left[
H^{(n)},D^{(n)}\right] \right) =  \label{projection} \\
P\odot \left( -\dsum\limits_{p}\delta _{p(n+1)}\underset{(n+1)}{\text{Tr}}%
\left( M^{(n+1\mid p)}\odot D^{(p)}\right) \right) \text{ ,}  \notag
\end{gather}%
and by using Eqs. (\ref{transitionmatrix})-(\ref{auxilary1}) we obtain 
\begin{gather}
\frac{\partial }{\partial t}D_{\alpha \alpha }^{(n)}+i\hbar ^{-1}\left(
H_{\alpha \gamma }^{(n)}D_{\gamma \alpha }^{(n)}-H_{\alpha \gamma
}^{(n)}D_{\gamma \alpha }^{(n)}\right) =  \label{hierarchy2} \\
\underset{(n+1)}{\text{Tr}}\left\{ D_{\gamma \delta }^{(n+1)}W_{\gamma
\delta ,\alpha \alpha }^{(n+1\mid n+1)}-\left( \sum_{\gamma }W_{\gamma
\gamma ,\alpha \alpha }^{(n+1\mid n+1)}\right) D_{\alpha \alpha
}^{(n+1)}\right\} \text{ .}  \notag
\end{gather}

For $n=1$ 
\begin{gather}
\frac{\partial }{\partial t}D_{\alpha \alpha }^{(1)}+i\hbar ^{-1}\left(
H_{\alpha \gamma }^{(1)}D_{\gamma \alpha }^{(1)}-H_{\alpha \gamma
}^{(1)}D_{\gamma \alpha }^{(1)}\right) =  \label{preboltzmann} \\
\underset{(2)}{\text{Tr}}\left\{ D_{\gamma \delta }^{(2)}W_{\gamma \delta
,\alpha \alpha }^{(2\mid 2)}-\left( \sum_{\gamma }W_{\gamma \gamma ,\alpha
\alpha }^{(2\mid 2)}\right) D_{\alpha \alpha }^{(2)}\right\} \text{ .} 
\notag
\end{gather}%
By neglecting the correlations which means assuming $D_{\alpha \beta
}^{(2)}(12)=D_{\alpha \beta }^{(1)}(1)D_{\alpha \beta }^{(1)}(2)$, we obtain%
\begin{align}
& \frac{\partial }{\partial t}D_{\alpha \alpha }^{(1)}(1)+i\hbar ^{-1}\left(
H_{\alpha \gamma }^{(1)}D_{\gamma \alpha }^{(1)}(1)-H_{\alpha \gamma
}^{(1)}D_{\gamma \alpha }^{(1)}(1)\right)  \notag \\
& =\underset{(2)}{\text{Tr}}\left\{ D_{\gamma \delta }^{(1)}(1)D_{\gamma
\delta }^{(1)}(2)W_{\gamma \delta ,\alpha \alpha }^{(2\mid 2)}(12)-\left(
\sum_{\gamma }W_{\gamma \gamma ,\alpha \alpha }^{(2\mid 2)}(12)\right)
D_{\alpha \alpha }^{(1)}(1)D_{\alpha \alpha }^{(1)}(2)\right\}
\label{boltzmann}
\end{align}%
which constitutes the quantum Boltzmann equation.

\section{Conclusions}

Here we have shown a representation of the statistical description of an
N-body system in terms of a density superoperator which reveals the
irreversibility at the macroscopic level. In the scenario drawn by the
Liouville-von Neumann equation with its associated Gibbs-von Neumann
relative entropy, irreversibility is hidden because of the inherent unitary
evolution of the N-particle density operator. Nonetheless, in the framework
of the quantum BBGKY-hierarchy the density superoperator follows a
non-unitary dynamics given by the generalized Liouville-von Neumann
equation. Thus, the generalization of the Gibbs-von Neumann entropy in terms
of the density supoeroperator we propose is not conserved by this dynamics.
Moreover, this dynamics is trace preserving which as is known \cite{lieb}-%
\cite{wherl}, ensures the non-negative character of the entropy production
which is coherent with the second law of Thermodynamic.

In addition, following the procedure of the non-Equilibrium Thermodynamics
we introduce a master matrix which satisfies a kind of Green-Kubo relation.
This, allows us to transform the BBGKY-hierarchy into a hierarchy of master
equations which we rewrite in terms of a transition superoperator related to
the master matrix. Once this has been done, by neglecting correlations in
the spirit of the Boltzmann's \textit{Stossahlansatz}, we derive the quantum
Boltzmann equation.

To summarize, our point is that the framework of the BBGKY-hierarchy which
explicitly incorporates the correlations between all the particle clusters
in the system constitutes the most appropriate description leading to
irreversibility at the macroscopic level. To manifest this irreversibility
all that is needed is a coherent definition of the entropy which is our
generalaized Gibbs-von Neumann relative entropy in addition to the
generalized Liouville-von Neumann equation.

In this way a macroscopic time arrow arises in a natural way in the system.
Hence, we can say that our analysis constitutes a microscopic basis of the
macroscopic irreversibility.

\end{document}